\documentclass[12pt]{article}
\usepackage{graphicx}

\newcommand\pubnumber{WSU--HEP--XXYY}
\newcommand\pubdate{\today}

\def\Title#1{\begin{center} {\Large #1 } \end{center}}
\def\Author#1{\begin{center}{ \sc #1} \end{center}}
\def\Address#1{\begin{center}{ \it #1} \end{center}}

\newcommand\pubblock{\rightline{\begin{tabular}{l} \pubnumber\\
         \pubdate  \end{tabular}}}
\newenvironment{Abstract}{\begin{quotation}  }{\end{quotation}}
\newenvironment{Presented}{\begin{quotation} \begin{center} 
             PRESENTED AT\end{center}\bigskip 
      \begin{center}\begin{large}}{\end{large}\end{center} \end{quotation}}
\def\Acknowledgements{\bigskip  \bigskip \begin{center} \begin{large}
             \bf ACKNOWLEDGEMENTS \end{large}\end{center}}

\def\beq{\begin{equation}}
\def\eeq#1{\label{#1}\end{equation}}
\def\eeqn{\end{equation}}

\def\beqa{\begin{eqnarray}}
\def\eeqa#1{\label{#1}\end{eqnarray}}
\def\eeqan{\end{eqnarray}}

\let\bar=\overbar

\def\Dslash{\not{\hbox{\kern-4pt $D$}}}
\def\dslash{\not{\hbox{\kern-2pt $\del$}}}

\def\msb{{\bar{\ssstyle M \kern -1pt S}}}

\begin{document}
\begin{titlepage}
\pubblock

\vfill
\Title{Leptonic Decays of Charm}
\vfill
\Author{Simon Eidelman}
\Address{Budker Institute of Nuclear Physics, Novosibirsk 630090, Russia}
\Address{Novosibirsk State University, Novosibirsk 630090, Russia}
\vfill
\begin{Abstract}
We present results of various searches for leptonic decays of charm mesons
performed with the Belle detector. Also discussed are $D^0 \to \gamma\gamma$ 
decays.
\end{Abstract}
\vfill
\begin{Presented}
The 7th International Workshop on Charm Physics (CHARM 2015)\\
Detroit, MI, 18-22 May, 2015
\end{Presented}
\vfill
\end{titlepage}
\def\thefootnote{\fnsymbol{footnote}}
\setcounter{footnote}{0}
%

\section{Introduction}

Leptonic decays of charm mesons are suppressed in Standard Model (SM) and
are therefore a convenient place to search for New Physics (NP). $B$ factories
provide a copious source of charm because
at $\Upsilon(4S)$ the cross section of $c\bar{c}$ production is $\sim 1.1$ nb,
so each fb$^{-1}$ brings $\sim 10^6$ events and with their huge integrated 
luminosities the $B$ factories produce inclusively a huge amount of
$c\bar{c}$ pairs. At lower energy, $\psi(3770)$ is a factory of 
$D^+D^-,~D^0\bar{D}^0$ while at $\sqrt{s} \sim 4.17$ GeV $D^+_sD^-_s$ pairs
are copiously produced.



\begin{figure}[htb]
\begin{center}
\hspace*{0.1cm} $c$ \hspace*{4cm} $\nu_{\ell}$ \\  
\vspace*{0.2mm}
{}

\includegraphics[width=0.6\textwidth]{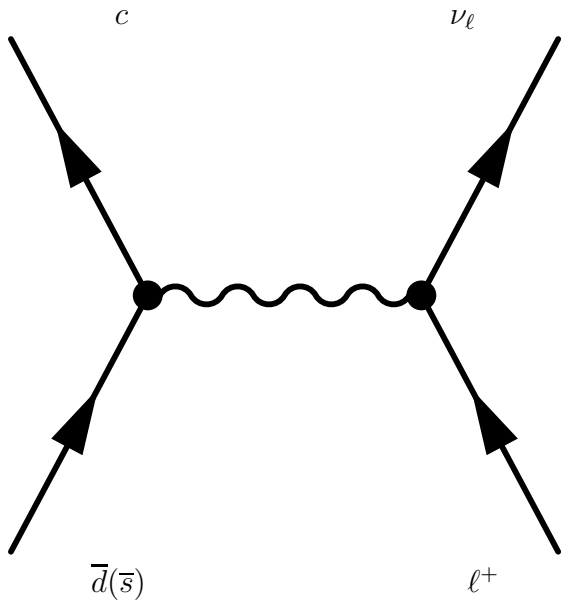} \\
\vspace*{0.2mm}
{}

\hspace*{0.1cm} $\bar{d}(\bar{s})$ \hspace*{4cm} $\ell^+$
\caption{Feynman diagram for $D^+_s$ leptonic decay.}
\label{fig:feynman}

\end{center}
\end{figure}

The branching fraction of $D_s$ leptonic decay (see the Feynman diagram in 
Fig.~\ref{fig:feynman}) is given by the following expression
 
$$
 {\cal B}(D_s^+ \to \ell^+\nu_{\ell})=\frac{\tau_{D_s}m_{D_{s}}}{8\pi}f_{D_{s}}^2
G_F^2|V_{cs}|^2m_{\ell}^2\left(1-\frac{m_{\ell}^2}{m_{D_{s}}^2} \right)^2.
$$
Here, $m_{D_{s}}$ is the $D_{s}^+$ meson mass, $\tau_{D_s}$ is its lifetime, 
$m_{\ell}$ is the lepton mass, $V_{cs}$ is the relevant
CKM matrix element, and $G_F$ is the Fermi coupling constant. 
The parameter $f_{D_{s}}$ is the $D_s$ meson decay constant related to the 
wave-function overlap of the meson's constituent quark and antiquark. 
The leptonic decays of pseudoscalar mesons 
are helicity suppressed with 
$\Gamma(\ell^+\nu_\ell) \propto m^2_{\ell}$, so that
$\Gamma(e^+\nu_e) \ll \Gamma(\mu^+\nu_\mu) \ll \Gamma(\tau^+\nu_\tau)$, \\
$R^{D_s}_{\tau/\mu}\equiv {\cal B}(D_s^+ \to \tau^+\nu_{\tau})/
                      {\cal B}(D_s^+ \to \mu^+\nu_{\mu})=
m_{\tau}^2/m_{\mu}^2\cdot(1-m^2_{\tau}/m^2_{D_s})^2/(1-m^2_{\mu}/m^2_{D_s})^2
=9.762\pm0.031$, $R^{D^+}_{\tau/\mu}\equiv {\cal B}(D^+ \to \tau^+\nu_{\tau})/
                      {\cal B}(D^+ \to \mu^+\nu_{\mu})= 2.67 \pm 0.01$. 
A study of such decays not only tests SM and searches for NP, 
e.g., a charged Higgs, but also provides a test of lepton flavor universality 
in decays with $\mu$ and $\tau$ (decays to $e^+\nu_e$ are extremely rare and 
hardly observable).

\section{$D_s^+ \to \ell^+\nu_{\ell}$ at Belle}
Recently Belle studied  $D_s^+ \to \ell^+\nu_{\ell}$ with 913 fb$^{-1}$ at
$\Upsilon(4S)$ and $\Upsilon(5S)$~\cite{zupanc}.
The $e^+e^- \to c\bar{c}$ events that contain $D_s^+$ mesons
are reconstructed in two steps. First,
one of the two charm quarks that hadronizes into a $D_s^{*+}$ meson,
is searched for. Then the other, a tagging charm hadron, $D_{\rm tag}$ 
($D^0,~D^-,~\Lambda_c^-,~D^{*-},~D^{*0}$), is reconstructed.
The strangeness of the event is conserved by requiring an additional kaon, 
denoted $K_{\rm frag}$, to be produced in the fragmentation process;
$K_{\rm frag}$ is either $K^+$ or $K^0_S$. In events where $D_{\rm tag}$ is the 
tagging charm hadron, 
the baryon number of the event is conserved by requiring an antiproton. 
Since Belle collected data at energies well above the
$D{}^{(\ast)}_{\rm tag} K_{\rm frag} D_s^{\ast-}$ threshold, additional particles 
can be produced in the course of hadronization. 
These particles are denoted as $X_{\rm frag}$ and consist of an even number 
of kaons plus any number of pions or photons. In this measurement,
only pions are considered when reconstructing the fragmentation system.
The number of inclusively reconstructed $D_s^+$ mesons is
extracted from the distribution of events in the missing mass,
$M_{\rm miss}(D_{\rm tag} K_{\rm frag}X_{\rm frag}\gamma)$, recoiling against the 
$D_{\rm tag} K_{\rm frag}X_{\rm frag}\gamma$ system
\begin{equation}
M_{\rm miss}(D_{\rm tag} K_{\rm frag}X_{\rm frag}\gamma) = 
\sqrt{p_{\rm miss}(D_{\rm tag} K_{\rm frag}X_{\rm frag}\gamma)^2},
\end{equation}
where $p_{\rm miss}$ is the missing four-momentum in the event
\begin{equation}
p_{\rm miss}(D_{\rm tag} K_{\rm frag}X_{\rm frag}\gamma)  =  
p_{e^+} +  p_{e^-} - p_{D_{\rm tag}} - p_{K_{\rm frag}} - p_{X_{\rm frag}} - p_{\gamma}.
\end{equation}
Here,  $p_{D_{\rm tag}}$, 
$p_{K_{\rm frag}}$, $p_{X_{\rm frag}}$, and $p_{\gamma}$ are the measured 
four-momenta of the reconstructed $D_{\rm tag}$, 
strangeness-conserving kaon, fragmentation system and the photon from
$D_s^{*+} \to D_s^+\gamma$. Correctly reconstructed 
events produce a peak in the 
$M_{\rm miss}(D_{\rm tag} K_{\rm frag}X_{\rm frag})$ at the nominal $D_s^+$ meson mass.

18 modes of $D_{\rm tag}$ were considered in total.
Six modes of $D^0$ (the total branching of 38.4\%):
$K^-\pi^+,~K^-\pi^+\pi^0,~K^-\pi^+\pi^+\pi^-,~K^-\pi^+\pi^+\pi^-\pi^0$,	
$K^0_S\pi^+\pi^-,~K^0_S\pi^+\pi^-\pi^0$, 	 
six modes of $D^+$ (28.0\%): 
$K^-\pi^+\pi^+,~K^-\pi^+\pi^+\pi^0,~K^0_S\pi^+$,	
$K^0_S\pi^+\pi^0,~K^0_S\pi^+\pi^+\pi^-,~K^+K^-\pi^+$ 
and six modes of  $\Lambda_c^+$ (16.8\%):
$pK^-\pi^+,~pK^-\pi^+\pi^0,~pK^0_S,~\Lambda\pi^+$,
$\Lambda\pi^+\pi^0,~\Lambda\pi^+\pi^+\pi^-$	  
There were seven $X_{\rm frag}$ modes of pions only:
nothing, $\pi^{\pm},~\pi^0,~\pi^{\pm}\pi^0,~\pi^{\pm}\pi^{\mp}$,
$\pi^{\pm}\pi^{\mp}\pi^{\pm}$,\\ $\pi^{\pm}\pi^{\mp}\pi^0$. With these conditions 
$94360 \pm 1310$ events were selected.

\begin{figure}
\vspace*{-2cm}
\begin{center}
{\hspace*{-3.5cm}
\includegraphics[width=0.8\textwidth]{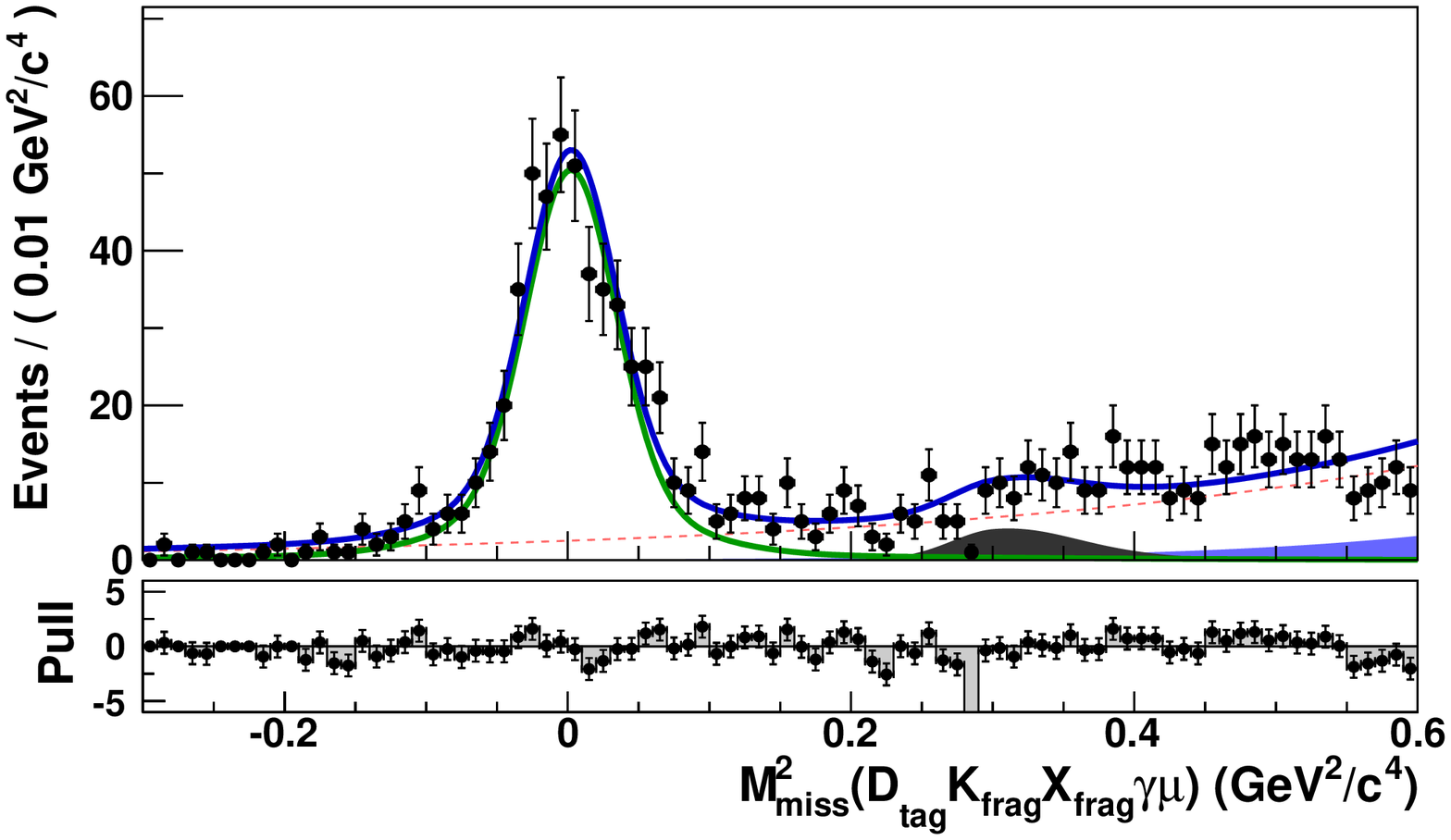}
\vspace*{-3.5cm}
\caption{Results of the fit for $D^+_s \to \mu^+\nu_{\mu}$}}
\label{fig:bel5}
\end{center}
\end{figure}

Results  of a search for $D^+_s \to \mu^+\nu_\mu$
are shown in Fig.~\ref{fig:bel5}.

\begin{figure}
\vspace*{-1.5cm}
\begin{center}
{\hspace*{-3.5cm}
\includegraphics[width=0.8\textwidth]{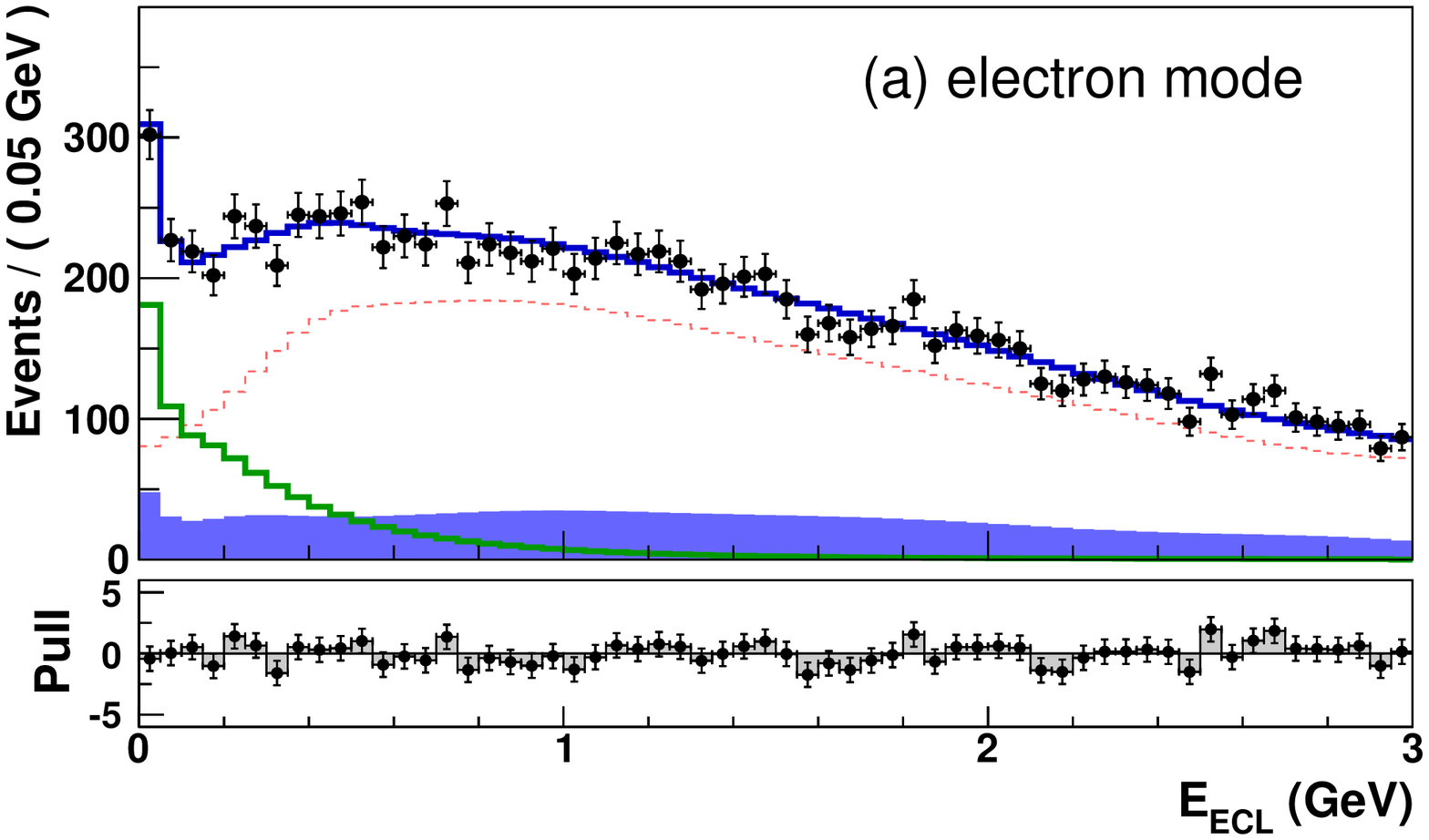}} \\
\vspace*{-2.5cm}
{\hspace*{-3.5cm}
\includegraphics[width=0.8\textwidth]{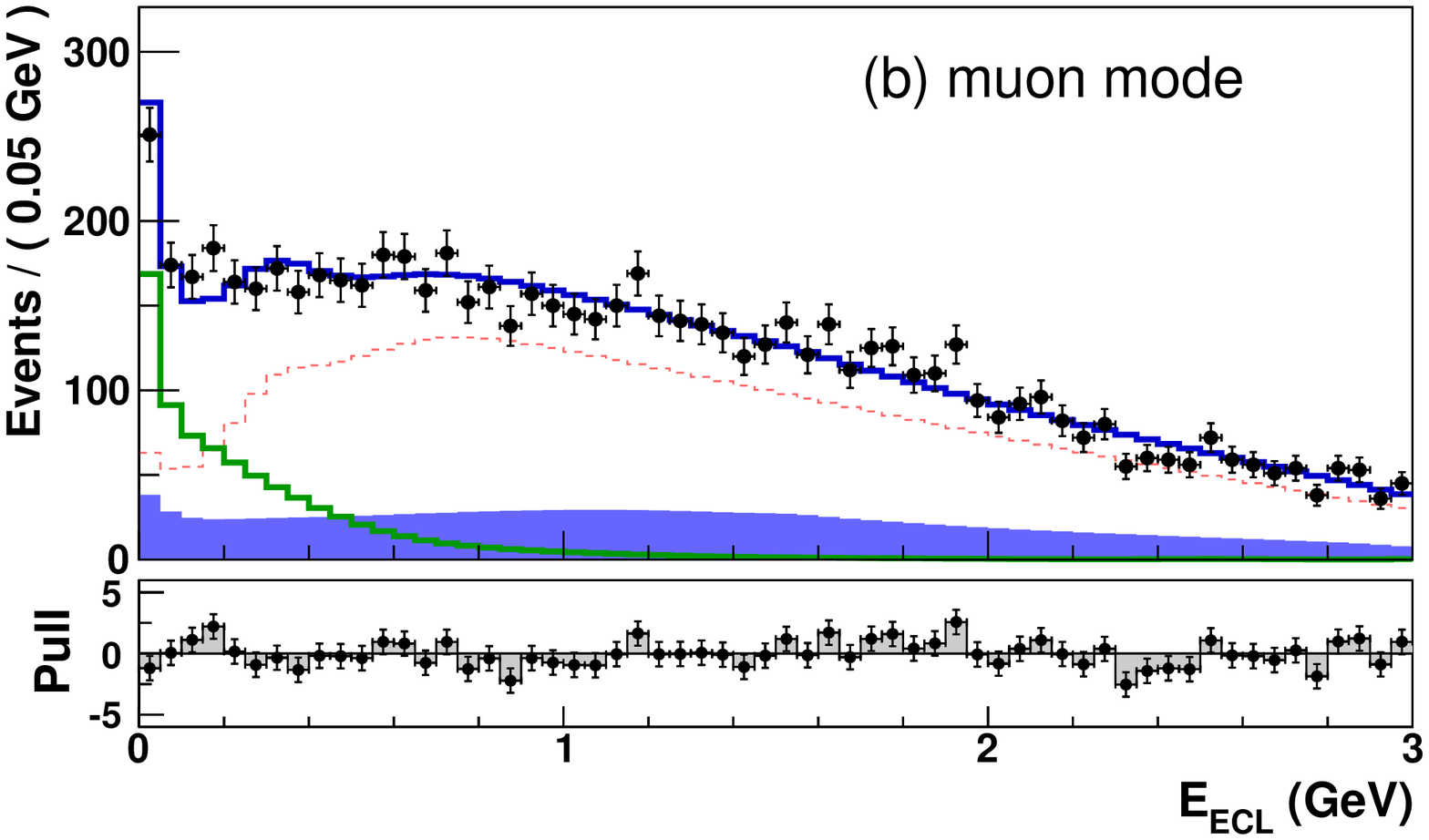}} \\
\vspace*{-2.5cm}
{\hspace*{-3.5cm}
\includegraphics[width=0.8\textwidth]{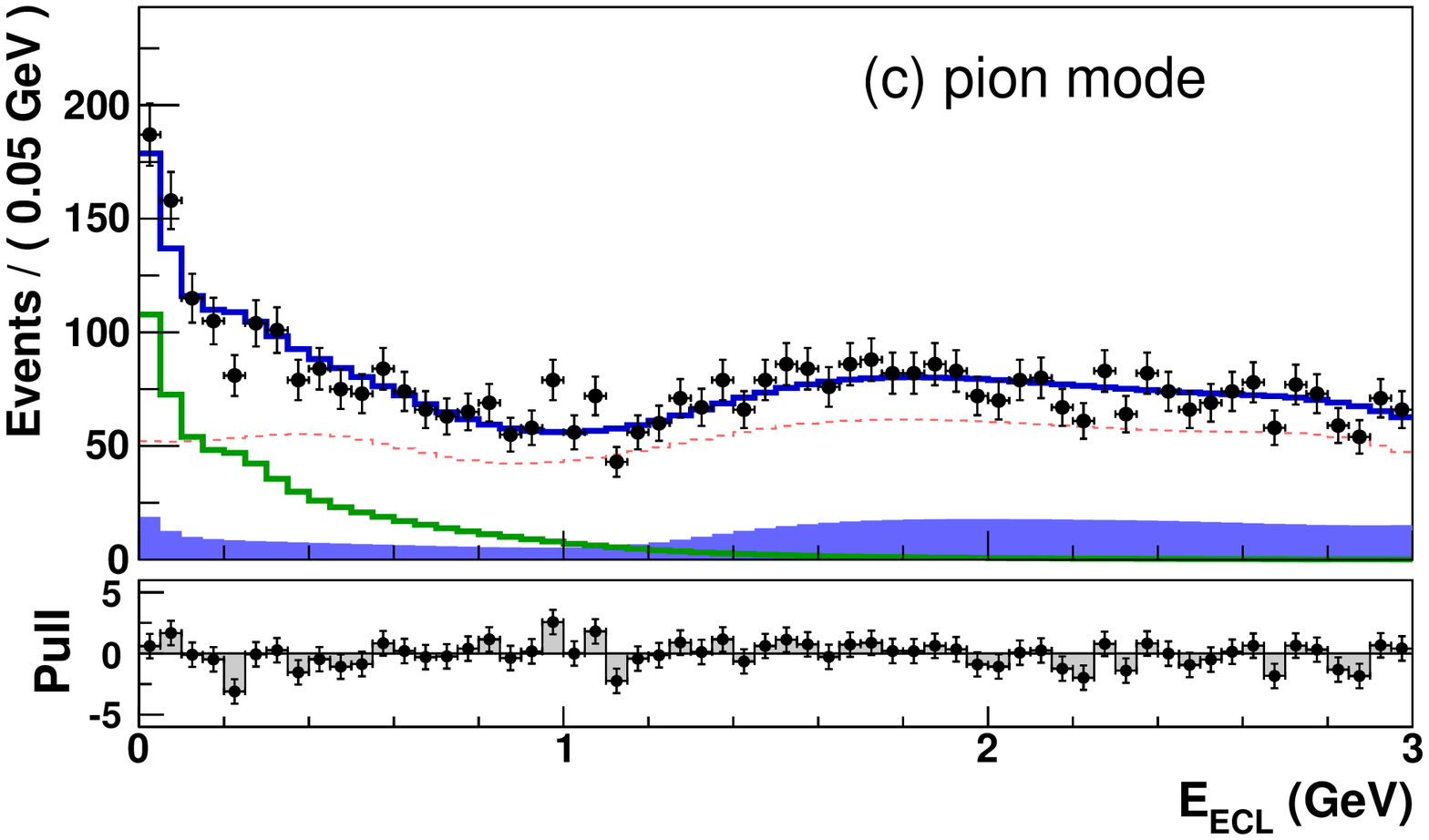}} \\
\vspace*{-1.5cm}
{\hspace*{-5cm}
\caption{Results of the fit for $D^+_s \to \tau^+ \nu_{\tau}$}}
\label{fig:bel6}
\end{center}
\end{figure}

In $D^+_s \to \tau^+\nu_\tau$ decay, because of extra $\nu$'s there is no peak 
in $M_{\rm miss}$, so small $E_{\rm ECL}$ is used instead, see Fig.~\ref{fig:bel6},
where three different decay modes of the $\tau^+$ are used.

Results of the fits are shown in Table~\ref{tab:1}. Since different $\tau$ 
decay modes give  consistent values of the branching fractions,  they are 
combined in the Table.

\begin{table}
\begin{center}
 \begin{tabular}{lccc}\hline
  $D_s^+$ decay mode	& Signal yield	& $f_{\rm bias}\cdot\varepsilon$ [\%]	& ${\cal B}$ [\%] \\
  \hline
  
  $\mu^+\nu_{\mu}$			& $\phantom{0}492\pm26\phantom{0}$	& $98.2$				& $0.531\pm0.028\pm0.020$\\
  $\tau^+\nu_{\tau}$ ($e$ mode)		& $\phantom{0}952\pm59\phantom{0}$	& $18.8$				& $5.37\pm0.33{}^{+0.35}_{-0.31}$\\
  $\tau^+\nu_{\tau}$ ($\mu$ mode)	& $\phantom{0}758\pm48\phantom{0}$	& $13.7$				& $5.86\pm0.37{}^{+0.34}_{-0.59}$\\
  $\tau^+\nu_{\tau}$ ($\pi$ mode)	& $\phantom{0}496\pm35\phantom{0}$	& $\phantom{0}8.7$			& $6.04\pm0.43{}^{+0.46}_{-0.40}$\\
  $\tau^+\nu_{\tau}$ (combined)		& $2217\pm83\phantom{0}$		& $41.2$				& $5.70\pm0.21{}^{+0.31}_{-0.30}$\\\hline
\end{tabular}
\caption{Results of data selection for the  $\mu^+\nu_{\mu}$
 and $\tau^+\nu_{\tau}$ decay modes}
\label{tab:1}
\end{center}
\end{table}

The obtained value of the branching fraction for the muon decay mode of $D_s$
\begin{equation}
 {\cal B}(D_s^+ \to \mu^+\nu_\mu)=  
(5.31 \pm 0.28({\rm stat.}) \pm 0.20({\rm syst.}))\times 10^{-3}
\end{equation}
 is consistent with and much more precise than the previous Belle 
one~\cite{belleold}:
\begin{equation} 
 {\cal B}(D_s^+ \to \mu^+\nu_\mu)=
(6.44 \pm 0.76({\rm stat.}) \pm 0.57({\rm syst.}))\times 10^{-3}.
\end{equation}
Comparison with measurements of the other groups is performed in
Table~\ref{tab:2}.    

\begin{table}
\begin{center}
\begin{tabular}{lcc}
\hline
 ${\cal B}(D_s^+ \to \mu^+\nu_{\mu}),~10^{-3}$ & $N_{\rm ev}$ & Group  \\
\hline
$5.31\pm 0.28 \pm 0.20$ & $492 \pm 26$ & Belle~\cite{zupanc} \\
$6.02\pm 0.38 \pm 0.34$ & $275 \pm 17$ & BaBar~\cite{babr1} \\
$5.65\pm 0.45 \pm 0.17$ & $235 \pm 14$ & CLEO~\cite{cleo1} \\
\hline
$5.56 \pm 0.25$ & -- & Average \\
\hline
\end{tabular}
\caption{Summary of $D_s^+ \to \mu^+\nu_{\mu}$ measurements}
\label{tab:2}
\end{center}
\end{table}

For the $\tau$ lepton decay mode Belle obtains
\begin{equation} 
{\cal B}(D_s^+ \to \tau^+\nu_{\tau}) = 
(5.70 \pm 0.21({\rm stat.}) {}^{+0.31}_{-0.30}({\rm syst.}))\times 10^{-2},
\end{equation}
doubling the total statistics of the previous experiments 
and consistent with the PDG2012 $(5.43 \pm 0.31) \times 10^{-2}$~\cite{pdg12}.
Comparison with other measurements is presented in Table~\ref{tab:3}.

\begin{table}
\begin{center}
\begin{tabular}{lcc}
\hline
 ${\cal B}(D_s^+ \to \tau^+\nu_{\tau}),~10^{-2}$ & $N_{\rm ev}$ & Group  \\
\hline
$5.70\pm 0.21^{+0.31}_{-0.30}$ & 2.2k & Belle~\cite{zupanc} \\
$5.00\pm 0.35 \pm 0.49$ & $748 \pm 53$ & BaBar~\cite{babr1} \\
$6.42\pm 0.81 \pm 0.18$ & $126 \pm 16$ & CLEO~\cite{cleo1} \\
$5.52\pm 0.57 \pm 0.21$ & $155 \pm 17$ & CLEO~\cite{cleo2} \\
$5.30\pm 0.47 \pm 0.22$ & $181 \pm 16$ & CLEO~\cite{cleo3} \\
\hline
$5.54 \pm 0.24$ & -- & Average \\
\hline
\end{tabular}
\caption{Summary of $D_s^+ \to \tau^+\nu_{\tau}$ measurements}  
\label{tab:3}
\end{center}
\end{table}


They also perform a test of lepton flavor universality, 
\begin{equation}
 R_{\tau/\mu}^{D_s} = 10.73\pm0.69({\rm stat.}){}^{+0.56}_{-0.53}({\rm syst.}),\\
\end{equation}
in agreement with the SM value of $9.762\pm0.031$.
 

As expected, a study of $D_s^+ \to e^+\nu_e$ decay does not show any signal
and they set an upper limit for the branching fraction 
\begin{equation}  
{\cal B}(D_s^+ \to e^+\nu_{e}) < 0.83 \times 10^{-4} ~{\rm at}~90\% CL 
\end{equation}
compared to the best previous limit $< 1.2 \times 10^{-4}$ from CLEO 
with 600 pb$^{-1}$ at 4.17 GeV~\cite{cleo1}.

The results for the branching fractions can be used for a determination 
of $f_{D_s}$ from the relation
\begin{equation}
f_{D_s}=\frac{1}{G_F m_{\ell}\left( 1-\frac{m_{\ell}^2}{m_{D_s}^2}\right)
\left|V_{cs}\right|} \sqrt{\frac{8\pi{\cal B}(D_s^+ \to\ell^+\nu_{\ell})}
{m_{D_s}\tau_{D_s}}}.
\end{equation}

From $|V_{ud}|=0.97425\pm0.00022$ and $|V_{cb}|=(40.9\pm1.1)\times10^{-3}$  
and using the relation $|V_{cs}|=|V_{ud}|-|V_{cb}|^2/2$ one obtains the 
following results, see Table~\ref{tab:4}.

\begin{table}[b]
\begin{center}
\begin{tabular}{lc}
\hline
  $D_s^+$ decay  & $f_{D_s}$ [MeV] \\ \hline
$\mu^+\nu_\mu$ & $249.8\pm6.6({\rm stat.})\pm4.7({\rm syst.})\pm1.7(\tau_{D_s})$ \\
$\tau^+\nu_\tau$ & $261.9\pm4.9({\rm stat.})\pm7.0({\rm syst.})\pm1.8(\tau_{D_s})$ \\\hline
  Combination       & $255.5\pm4.2({\rm stat.})\pm4.8({\rm syst.})\pm1.8(\tau_{D_s})$ \\\hline
\end{tabular}
\caption{Determination of $f_{D_s}$ at Belle}
\label{tab:4}
\end{center}
\end{table}

The combined result from the two decay modes is 
consistent with the most precise value from lattice QCD 
$248.0 \pm 2.5$ MeV~\cite{davies}.


\section{$D^0 \to \gamma\gamma$ and $D^0 \to \ell^+\ell^-$ Decays}

In SM, flavor-changing neutral current (FCNC) decays are suppressed by the
GIM mechanism, so that the expected branching fractions
for $D^0 \to \gamma\gamma$ decays are small
\begin{equation} 
{\cal B} \sim (1-3) \cdot 10^{-8}.
\end{equation}
In MSSM, gluino exchange enhances ${\cal B}$ to $6 \cdot 10^{-6}$,
therefore searches for NP can be performed. The current
sensitivity of such searches is at the level of a few units of
$10^{-6}$ and the achieved upper limits are shown in Table~\ref{tab:2g}. 

\begin{table}
\begin{center}
\begin{tabular}{lcc}
\hline
Group & $\int{Ldt},~{\rm fb}^{-1}$  & ${\cal B},~10^{-6}$  \\
\hline
CLEO~\cite{cleo5} & 13.8 & $<29$  \\
BaBar~\cite{babr2} & 470.5  &  $<2.2$ \\
BESIII~\cite{bes2} & 2.92 & $<3.8$  \\
Belle & 832  & In progress \\
\hline
\end{tabular}
\caption{Branching fractions of $D^0 \to \gamma\gamma$ decays}
\label{tab:2g}
\end{center}
\end{table}

In SM, the FCNC   $D^0 \to \ell^+\ell^-$   decays are additionally suppressed
by helicity,
\begin{equation}  
{\cal B} \sim 2.7 \cdot 10^{-5} {\cal B}(D^0 \to \gamma\gamma)
\end{equation}
or $\sim 10^{-13}$. Extensions of SM, e.g., models with R-parity violating 
SUSY, large extra dimensions or leptoquarks enhance the branchings to 
$\sim 10^{-8}$. Even smaller in SM are expected branchings for 
lepton-flavor-violating decays. The results of searches for such 
decays are presented in Table~\ref{tab:ll}.

\begin{table}[h]
\begin{center}
\begin{tabular}{llcc}
\hline
Mode & Group & $\int{Ldt},~{\rm fb}^{-1}$  & ${\cal B}$ \\
\hline
$D^0 \to e^+e^-$ & Belle~\cite{petric} & 660  & $<7.9 \cdot 10^{-8}$  \\
$D^0 \to \mu^+\mu^-$ & LHCb~\cite{lhcb} & 0.9  &  $<6.2 \cdot 10^{-9}$  \\
$D^0 \to e^\pm \mu^\mp$ & Belle~\cite{petric} & 660 &  $<2.6 \cdot 10^{-7}$  \\
\hline
\end{tabular}
\caption{Branching fractions of  $D^0 \to \ell^+\ell^-$ decays}
\label{tab:ll}
\end{center}
\end{table}

\section{Conclusions}
Leptonic decays of $D^0,~D^+,~D^+_s$ are very convenient
to search for effects of New Physics. Recently there has been significant 
experimental progress due to CLEO, BESIII, BaBar, Belle and LHCb.
LHCb has strong advantage for decays accessible to it because of large 
data samples at high energy. Experience of CLEOc and BESIII shows 
advantages of exclusive measurements with 
$e^+e^- \to D^+D^-,~D^0\bar{D}^0,~D^+_sD^-_s$.
Future progress is related to LHCb, BelleII and hopefully Super-c-$\tau$.

\section{Acknowledgments}
I'm grateful to the organizers and in particular to Alexey Petrov for
the excellent Conference. This work was supported in part by the
RFBR grants 13-02-00215, 14-02-91332, 15-02-05674 and the DFG grant
HA 1457/9-1.

\end{document}